\documentclass[pre,longbibliography,showpacs,showkeys,twocolumn,superscriptaddress,
              floats,floatfix,preprintnumbers]{revtex4-1}
\usepackage{amssymb,amsmath}
\usepackage{bm,url,graphics,overpic,subfigure,epsfig,rotating,float}
\usepackage{mathrsfs}
\usepackage{color}

\newcommand{\ma}[1]{\ensuremath{\mathbb{#1}}}

\def \vvt {\bm{v}_{2}}

\def \Dtwo {D_{2}}
\def \muc {\mu_{\rm c}}
\def \zast {z^{\ast}}

\def \mP {\mathscr{P}}

\def \at {a_{t}}

\def \uu  {{\bm u}}
\def \vv  {{\bm v}}
\def \ff  {{\bm f}}

\def \ueta {u_{\rm \eta}}
\def \teta {\tau_{\rm \eta}}
\def  \xx  {{\bm x}}

\def  \VV  {{\bm V}}

\def \p {p}
\def \cs {c_{\rm s}}
\def \Lx {L_x}
\def \Ly {L_y}
\def \Lz {L_z}

\def  \RR  {{\bm R}}
\def  \Vc  {V_{\rm c}}
\def  \Rc  {R_{\rm c}}
\def  \VR  {V_{R}}
\def  \VRa  {|V_{R}|}

\def \taup {\tau}

\def \curl {{\bm \nabla} \times}

\def \delt {\partial_t}

\newcommand{\bra}[1]{\left\langle #1\right\rangle}
\def \Rey  {\mbox{Re}}
\def \Ma  {\mbox{Ma}}
\def \ao  {a_1}
\def \at  {a_2}
\def \St  {\mbox{St}}
\def \Sto  {\mbox{St}_{\rm 1}}
\def \Stt  {\mbox{St}_{\rm 2}}
\def \Stm  {\overline{\mbox{St}}\protect\phantom{|}}

\def \Teddy {T_{\rm eddy}}

\def \kf  {k_{\rm f}}

\def \urms  {u_{\rm rms}}

\def \Np  {N_{\rm p}}

\def \Rc {R_{\rm c}}

\newcommand{\DDt}[1]{\frac{{\rm D} #1}{{\rm D}t}}
\newcommand{\eq}[1]{~(\ref{#1})}
\newcommand{\Eq}[1]{Eq.~(\ref{#1})}
\newcommand{\Fig}[1]{Fig.~(\ref{#1})}

\newcommand{\bfig}{\begin{figure}}
\newcommand{\efig}{\end{figure}}
\newcommand{\bc}{\begin{center}}
\newcommand{\ec}{\end{center}}
\newcommand{\bea}{\begin{eqnarray}}
\newcommand{\eea}{\end{eqnarray}}

\begin{document}
\title{Relative velocities  in bi-disperse turbulent aerosols:
simulations and theory}
\author{Akshay Bhatnagar}
\email{akshayphy@gmail.com}
\affiliation{ Nordita, KTH Royal Institute of Technology and
Stockholm University, Roslagstullsbacken 23, 10691 Stockholm, Sweden}
\author{K. Gustavsson}
\affiliation{Department of Physics, Gothenburg University, 41296 Gothenburg, Sweden}
\author{B. Mehlig}
\affiliation{Department of Physics, Gothenburg University, 41296 Gothenburg, Sweden}
\author{Dhrubaditya Mitra}
\email{dhruba.mitra@gmail.com}
\affiliation{ Nordita, KTH Royal Institute of Technology and
Stockholm University, Roslagstullsbacken 23, 10691 Stockholm, Sweden}
\preprint{2018-088}
\begin{abstract}
We perform direct numerical simulations of a bi-disperse suspension
of heavy spherical particles in forced, homogeneous, and isotropic three-dimensional
turbulence. We compute the joint distribution
of relative particle distances and longitudinal relative velocities 
between particles of different inertia.
 For a pair of particles with small difference in their inertias we  
  compare our results with recent theoretical predictions
[Meibohm {\it et al.} Phys. Rev. E {\bf 96} (2017) 061102] for the shape of this distribution.
We also compute the moments of relative velocities  as a function of particle separation, and 
compare with the theoretical predictions. We observe good agreement. For a pair of
particles that are very different from each other -- one is heavy and the other one has
negligible inertia -- we give a new theory to calculate their root-mean-square relative
velocity. This theory also agrees well with the results of our simulations.  
\end{abstract}

\maketitle

\section{Introduction}
Here we are concerned with  small but heavy particles moving  in a turbulent
flow. How frequently and at what speeds do such
particles collide with each other in turbulence?
This question plays a central role in attempting to understand collisions and coalescence of microscopic water droplets in turbulent  clouds~\cite{Pruppacher2010microphysics}, and to understand the  formation of planetesimals in proto-planetary disks~\cite{Wil08,Arm10,Anders}.
The particles in these turbulent aerosols are small and collisions between them
are few and far between, consequently fluctuations matter.
To understand how the distribution of particle sizes changes as a function of time,
it is therefore not sufficient to merely consider the average collision rate. To account for the fluctuations
it is necessary to consider the joint distribution of particle separations and their relative velocities \cite{Wei93,gustavsson2014relative,Win12}. A mean-field like description based solely on the first moment
of relative particle velocities neglects fluctuations and may therefore not be reliable.

V\"o{}lk {\it et al.}~\cite{volk1980collisions,Mizuno88,mar+miz+volk91} and others  \cite{meh+usk+wil07,Gus08b} formulated {\em inertial-range} theories for relative velocities of particles, referring to particle separations in the inertial range of turbulence. A criticism of this approach is that the collisions between the particles happen 
deep inside the dissipation range when the particle sizes are much smaller
than the Kolmogorov length, $\eta$.  It has been observed in direct numerical simulations (DNS) that inertial-range theories for the moments of relative velocities \cite{volk1980collisions,Mizuno88,mar+miz+volk91}  fail at small Stokes numbers \cite{ishihara18} (the Stokes number is a dimensionless measure of the importance of particle inertia). 
The predictions of Ref.~\cite{Gus08b} for the far tail of the distribution of relative velocities between nearby identical particles assume large Stokes numbers and a well-developed inertial range. This is difficult to achieve in DNS, and therefore it remains to be determined under which circumstances the prediction 
may hold.

Gustavsson {\it et al.} \cite{gustavsson2011distribution,Gus12,gustavsson2014relative,gustavsson2016collisions} developed a {\em dissipation-range} theory for the distribution of relative velocities of identical particles, when the collision radius -- the sum of the particle radii -- is in the  dissipation range of turbulence.
An asymptotic form of the distribution was obtained by matching two limiting cases and using that inertial particles of identical sizes distribute on a fractal attractor in phase space~\cite{gustavsson2011distribution,gustavsson2014relative}.  The result is a non-Gaussian distribution, with power-law  tails that reflect large fluctuations. The theory applies in the limit where the Stokes number is large enough for particles to detach from the streamlines of the flow. But since the theory \cite{gustavsson2011distribution,Gus12,gustavsson2014relative,gustavsson2016collisions} neglects inertial-range fluctuations, it may require modifications at very large Stokes numbers where the particle separations  explore the
inertial range. 

In the astrophysical literature, DNS results for the relative-velocity distribution were recently reported by 
Ishihara {\em et al.} \cite{ishihara18}, as well as by Pan and Padoan
\cite{pan2013turbulence,pan2014distri}. These authors
fit the distribution to stretched exponentials. This raises the question how universal  the power-law
tails predicted in Refs.~ \cite{gustavsson2011distribution,gustavsson2014relative} are. For 
Stokes numbers of order unity, the power laws were clearly seen in 
DNS~\cite{perrin2015relative,bhatnagar2018statistics}.

The findings  and open questions described above apply to identical
particles. But to understand how the size distribution of particles in
turbulent aerosols changes as a result of collisions and coalescences,
the distribution for particles of different sizes (different Stokes
numbers) is needed. Meibohm {\it et al.} \cite{meibohm2017relative}
developed  a {\em dissipation-range} theory for the distribution of relative velocities
of particles that have different Stokes numbers, by analyzing a statistical model in the white-noise limit.  The predictions of Ref.~\cite{meibohm2017relative}  have not been tested in DNS yet. 

To understand the distribution of relative velocities in turbulent aerosols
is an important problem to study -- both in theory and in simulations -- because it is hard to obtain direct measurements of droplet velocities in clouds, and quite impossible as far as grain velocities in proto-planetary disks are concerned. There are two laboratory experiments~\cite{Saw14,dou+bra+ham+etal18}
that have measured the distribution of relative velocities of micron-sized particles in turbulence, and
 their mean and root-mean square values as functions of particle separations. 
  Experimental limitations make it difficult to measure at which relative velocities particles actually collide in these experiments.
 For micron-sized particles this occurs at separations deep inside the dissipative range, at present outside the spatial resolution of the experiments.

It is therefore important to validate existing theories for collision velocities of particles in turbulence by comparison
 with results of DNS. This is the purpose of the present paper. 
It is organized as follows: in Section~\ref{model}
we describe the model and details of the DNS.  
In Section~\ref{sec:theory} we summarize
the key theoretical results of Refs.~\cite{gustavsson2011distribution,meibohm2017relative}. 
In Section~\ref{results} we present our DNS results for the 
relative velocities between particles with different Stokes numbers.  We  compare the DNS results
for the  joint probability distribution of relative velocities and separations with
the theoretical predictions of Meibohm {\it et al.} \cite{meibohm2017relative}. The distribution is non-Gaussian. When the difference between the Stokes numbers is not too large, then the distribution exhibits power-law tails
as predicted by theory. At small separations and relative velocities, the power law in relative velocities is cut off,
it becomes a broad Gaussian (approximately uniform), verifying the new velocity scale $\Vc$ predicted by theory \cite{meibohm2017relative}. Also the distribution
of separations becomes uniform for separations smaller than $\Rc$. This scale was predicted in 
Refs.~\cite{Chu05,Bec05}. We show how the scales $\Vc$ and $\Rc$ are related. 
Finally, we develop a new theory for the root-mean-square (RMS) relative velocities of particles
when one of the particles has very small Stokes number. We find that the results from this theory are in
accord with our simulations. We conclude in Section~\ref{discussion}. 

\section{Numerical method}
\label{model}
\subsection{Particle dynamics}
We describe the  motion of a heavy particle in a turbulent flow by the
Stokes model \cite{gus+meh16}:
\begin{align}
\tfrac{{\rm d}}{{\rm d}t} \xx &= \vv \/ \label{eq:dxdt}\,,\quad
\tfrac{{\rm d}}{{\rm d}t}\vv = \frac{1}{\taup}\left[ \uu({\xx,t}) - \vv \right] \,.
\end{align}
Here $\xx$ and $\vv$ are
 the position and velocity of the particle,
the characteristic response time of the particle is $\taup$. The response time depends
upon the particle size, $a$. In the Stokes limit, $\tau=(2\rho_p/9\rho)\, a^2/\nu$. Here
$\rho_p$ and $\rho$ are the mass densities of the particle and the fluid,
and $\nu$ is the kinematic viscosity.
Finally $\uu(\xx,t)$ is the flow velocity.
This model assumes that the effect of gravitational
acceleration is small compared to the acceleration due to the
turbulent flow,  fluid-inertia corrections are small, and 
both particle-particle interactions and Brownian diffusion of individual particles
are ignored.

\subsection{Direct numerical simulation of turbulence}
The flow velocity $\uu(\xx,t)$ is determined by solving the
Navier--Stokes equation
\begin{subequations}
\begin{align}
\tfrac{\partial }{\partial t}\rho &+ \nabla \cdot (\rho \uu) = 0  \/\,, \label{eq:density}\\
\rho \tfrac{{\rm D} }{{\rm D}t}\uu &= -\nabla \p + \mu \nabla \cdot {\ma S} + {\bm f}  \,.
\label{eq:mom}
\end{align}
\label{eq:fluid}
\end{subequations}
Here $\tfrac{\rm D}{{\rm D}t} \equiv \delt + \uu \cdot \nabla$ is the Lagrangian
derivative,  $\p$ is the pressure of the fluid,
and $\rho$ is its density as mentioned above.
The dynamic viscosity is denoted by $\mu\equiv\rho\nu$, and
${\ma S}$  is the second-rank tensor with components
$S_{kj} \equiv \partial_k u_j + \partial_j u_k -\delta_{jk}(2/3)\partial_l u_l$
(Einstein summation convention).
Here $\partial_k u_j $ are the elements of the matrix $\ma A$ of fluid-velocity gradients.
We use the ideal gas equation of state with a constant
speed of sound.  

Our simulations are performed in a three-dimensional periodic box
with sides $\Lx=\Ly=\Lz=2\pi$ in code units.
To solve Eqs.~(\ref{eq:fluid}) we use the pencil code~\cite{pencil-code},
which uses a sixth-order finite-difference scheme for space
derivatives and a third-order Williamson-Runge-Kutta~\cite{wil80} scheme for time
derivatives.
The external force $\ff$, which is a white-in-time, Gaussian, stochastic
process concentrated on a shell of wavenumber with
radius $\kf$ in Fourier space~\cite{B01}, is 
 integrated by using the Euler--Marayuma scheme~\citep{hig01}.  
Under the action of the force the flow attains a statistically stationary state where
the average energy dissipation by viscous forces is balanced by
the average energy injection by the external force, $\ff$.
The amplitude of the external force is chosen such that the 
Mach number, $\Ma \equiv \urms/\cs$ is always less than $0.1$, i.e.,
the flow is weakly compressible which has no important effect on our
results; please see the discussion in Ref.~\cite{bhatnagar2018statistics}, section II and
Appendix A in Ref~\cite{bhatnagar2018statistics} for further details. 
The same setup has been used before in
studies of scaling and intermittency in fluid and magnetohydrodynamic
turbulence~\cite{dob+hau+you+bra03,hau+bra+dob03,hau+bra04}.

We introduce the particles into the simulation
after the flow has reached a statistically stationary state.
Initially, the positions of the heavy particles are random and
statistically homogeneous with zero initial velocity. 
Then we simultaneously solve Eqs.~(\ref{eq:dxdt})  and \eq{eq:fluid}.
To this end we must interpolate the flow velocity to
typically off-grid positions of the heavy inertial particles.
We use a tri-linear method for interpolation.

\begin{table}
\begin{center}
\caption{Parameters for our DNS runs with $N^3$ collocation points:  
$\nu$ is the kinematic viscosity, 
$\Np$ is the number of particles,
$\Rey \equiv \urms/(\nu\kf)$ is based on the forcing wavenumber $\kf$, 
$\varepsilon$ is the mean rate of energy dissipation, 
$\eta \equiv (\nu^3/\epsilon)^{1/4}$, and 
$\teta \equiv (\nu/\epsilon)^{1/4}$ are the Kolmogorov length and time scales respectively,
and $\Teddy \equiv 1/(\urms\kf)$ is the large-eddy-turnover time.
The Mach number $\Ma = \urms/\cs \approx 0.1$.  
In the table we quoted dimensionless numbers.}

\begin{tabular}{c c c c c c c c}
\hline\hline
$N$ &  $\Np$ & $\Rey$ & $1/(\kf\eta)$ &  $\Teddy/\teta$ \\
\hline
$512$ & $10^7$ & $89$ & $14.28$ & $2.21$ \\
\hline
\end{tabular}
\end{center}
\label{table:para}
\end{table}
We define the Reynolds number by $\Rey\equiv \urms/(\nu\kf)$,
where $\urms$ is the root-mean-square velocity of the flow
averaged over the whole domain and the kinematic
viscosity $\nu$ 
The mean energy dissipation rate
$\varepsilon \equiv 2\nu \Omega$
where the enstrophy 
$\Omega \equiv \bra{\omega^2}$, and 
$\omega \equiv \curl \uu$ is the vorticity.  
The Kolmogorov length  is 
defined as
$\eta \equiv (\nu^3/\varepsilon)^{1/4}$,
the characteristic time scale of dissipation is given by  
$\teta = (\nu/\varepsilon)^{1/2}$
and 
$\ueta \equiv \eta/\teta$ is the characteristic velocity scale
at the dissipation length scale. 
In what follows, unless otherwise stated, we use
$\eta$, $\teta$, and $\ueta$ to non-dimensionalize 
length, time, and velocity respectively.  
The large eddy turnover-time is given by 
$\Teddy\equiv 1/(\kf\urms)$.
We define the Stokes number as $\St \equiv \taup/\teta$,
where $\tau$ is the particle response time in Eq.~(\ref{eq:dxdt}).
As mentioned in the Introduction, this parameter measures the importance of particle inertia. 

It is important to note that the particles in 
our simulations are
actually point particles. As particle-particle interactions are ignored
there are no real collisions.  As far as the numerical code
is concerned, the particles are characterized by the time-scale $\taup$
which determines the Stokes number. To estimate the radius of a particle
from its Stokes number we have used typical values of the ratio of the
density of the particle to the density of the background fluid that
corresponds to water droplets in clouds~\cite{sha03}. To obtain collision velocities
that corresponds to dust in proto-planetary disks one must use
different value of the density ratio. Also, since the sizes of the dust grains are smaller than the
mean-free-path of the gas~\cite{Eps24,Wil08,Arm10}, we must use a different expression
for the particle response time. It is obtained by replacing the mean free path $\ell$ in
$\nu = \ell c_{\rm s}$ (where $c_{\rm   s}$ is the sound speed) by the particle size $a$.
This yields $\tau \sim a$ instead of the quadratic dependence $\tau \sim a^2$ in Stokes law.

\section{Theoretical background}
\label{sec:theory}

In this Section we summarize the dissipation-range theory for the distribution of relative velocities between two particles
with different Stokes numbers \cite{meibohm2017relative}. We denote the relative-particle velocity by
 $\VV = \vv_2-\vv_1$, where  $\vv_1$ and $\vv_2$ are the individual particle velocities.
The distance between the particles is denoted by $R=|\RR|$, where $\RR=\xx_2-\xx_1$ is the
separation vector between the particle positions, and the longitudinal relative velocity
is defined as $\VR = \VV\cdot\RR/R$.
We denote the steady-state distribution of relative velocities and separations by $\mP(R,\VR)$. The moments of the distribution are characterized by
\begin{equation}
 \langle \VRa^p\rangle\! \equiv\! \frac{m_p(R)}{m_0(R)}\,,
\label{eq:mp}
m_p(R) \!= \!\!\int\!\! {\rm d}\VR\, \VRa^p \,\mP(R,\VR)\,.
\end{equation}
The factor $m_0(R)$ is related
to the pair correlation function by $m_0(R) \propto g(R) R^{d-1}$ \cite{gustavsson2014relative}.

\subsection{Distribution of relative velocities and separations}
Gustavsson and Mehlig \cite{gustavsson2011distribution,Gus12,gustavsson2014relative} developed a theory for the distribution of relative velocities of nearby {\it identical} particles. The theory  takes into account particle inertia, and it
  rests on the observation that such particles form fractal spatial patterns in turbulence \cite{gus+meh16}, and
that caustics can give rise to large relative velocities at small separations \cite{falkovich2002acceleration,wilkinson2005caustics,wilkinson2006caustic}.
 The theory predicts that the distribution of relative velocities $\VR$ at small separations  $R$ is a power law, reflecting fractal clustering in phase space.
The power-law exponent is related to   the  phase-space correlation dimension $D_2$   \cite{gustavsson2011distribution,gustavsson2014relative,meibohm2017relative}.
The  distribution determines the scaling of relative-velocity moments (\ref{eq:mp}) with separation $R$ \cite{Gus12}.
These predictions for {\it identical} particles should hold for turbulence as well as statistical-model flows.
In the white-noise limit, the theory was derived from first principles in Refs.~\cite{gustavsson2011distribution,gustavsson2014relative}.
 For turbulent flows,  the theoretical predictions were verified using
DNS~\cite{Vos13,perrin2015relative,bhatnagar2018statistics}
 and using  kinematic turbulence simulations \cite{Gus12}. See also Refs.~\cite{Bec10,Cen11,Sal12,jam+ray18}.

The correlation dimension $\Dtwo$ is not universal. In the white-noise limit   $\Dtwo$  can be calculated  in perturbation theory \cite{gus+meh16,gustavsson2011distribution}, but in general it must be determined numerically. As is well known, $\Dtwo$ depends non-monotonically on $\St$ with a minimum at $\St$ of order unity~\cite{bec2007heavy}.

Particles with {\it different} Stokes numbers
cluster on distinct fractal attractors, so that the distribution of separations between particles
with different Stokes numbers is cut off at a small spatial scale, $\Rc$ that depends on the difference
between the Stokes numbers  \cite{Chu05,Bec05}. How are the relative velocities of nearby particles affected?
In Ref.~\cite{meibohm2017relative} a statistical model for relative velocities between
particles with different Stokes numbers was analyzed in the white-noise limit. It was
shown that there is a new velocity scale $\Vc$, and that the distribution of $\VR$ and $R$ is a broad Gaussian below these scales \cite{meibohm2017relative},
in other words approximately uniform:
\begin{widetext}
\begin{align}
\mP(R,\VR)
= \mathscr{N}  R^{d-1}
\begin{cases}
1 &\text{for $|\VR |< \Vc$ and  $R < {\Vc/\zast} $\,,} \\
R^{\muc-d-1} &\text{for $R > {\Vc/\zast}$ and $\VRa < \zast R$\,,  } \\
\big({\VRa}/{\zast}\big)^{\muc-d-1} & \text{for $ \VRa > \Vc $ and $\zast R< |\VR|$ \,,}\\
0 & \text{for $|\VR| > V_0$\,.}
\end{cases}
\label{eq:muc}
\end{align}
\end{widetext}
In addition to the normalization  $\mathscr{N}$ there are four more parameters 
in Eq.~(\ref{eq:muc}):  the two velocity scales
$V_{\rm c}$ and $V_0$, the power-law exponent $\mu_{\rm c}$, and the parameter $z^\ast$.

The last parameter, $\zast$, defines the line
$|\VR| = z^\ast R$
in the $R$-$\VR$ plane where known
limiting behaviors of $\mP(R,\VR)$ in the dissipative range are matched
to obtain the theoretical predictions for $\mP(R,\VR)$.

The exponent $\muc$ is related to the phase-space correlation dimension $D_2(\Stm)$
of the mono-disperse system with Stokes number $\Stm$
 \begin{equation}
 \label{eq:muc2}
 \muc = \textrm{min}\{D_2(\Stm),d+1\}\,,
 \end{equation}
where $d=3$ is the spatial dimension, and $\Stm$ is the harmonic mean of the two Stokes numbers,
\begin{equation}
\Stm = \frac{2\St_1\St_2}{\St_1+\St_2}\,.
\end{equation}
The parameter $D_2$ can be calculated analytically in the white-noise limit \cite{Wil10b,Wil14,meibohm2017relative}, but in turbulent flows it must be determined numerically. 

Now consider the upper velocity scale $V_0$. 
It was  assumed in deriving Eq.~(\ref{eq:muc}) that it suffices  to consider separations in the dissipative range where the turbulent fluid velocities are spatially smooth. This range extends up to separations $R$ somewhat larger than the Kolmogorov length $\eta$. The theory mirrors the distribution of spatial separations for $R< 1$ to distributions
in relative velocities, just as it does for identical particles. Therefore the upper
cutoff for the $\VR$ power laws is
\begin{equation}
\label{eq:v0}
V_0=\zast\,.
\end{equation}
 How this parameter
depends upon the Stokes number is not known in general. In a one-dimensional statistical model
this parameter was calculated in the white-noise limit in Ref.~\cite{gustavsson2014relative}.

In Eq.~(\ref{eq:v0}), the distribution was simply set to zero for $\VR>V_0$. 
This  is an oversimplification, in particular for turbulence where  the far tails of the
$\VR$-distribution at small spatial separations result from particle pairs that have had separations in the inertial range in the past. For large
Stokes numbers and when the inertial range is well developed  it was argued  in Ref.~\cite{Gus08b} that 
the tail of the conditional distribution 
$\mP(R\!=\!0,\VR)$ has the form $ \sim {C_1}/(\varepsilon\tau)^{1/2}\, \exp[-C_2 |\VR|^{4/3}/(\varepsilon \tau)^{2/3}]$
for very large Stokes numbers. A statistical-model calculation with an inertial range yields the prefactors $C_1$  and $C_2$ in the white-noise limit, but they could have different parameter dependencies in turbulence \cite{Gus13a}. 
At smaller $\Rey$, when the inertial range is not well developed, one may argue that the tail
should be well approximated by a Gaussian with variance $\propto u_{\rm rms}^2$. The RMS turbulent velocity is 
an estimate
of the relative velocities of particles that move independently at large separations, of the order of the system size.
In summary, the far tail of the relative-velocity distribution is not universal. Here we simply set
\begin{equation}
V_0 = u_{\rm rms}
\end{equation}
when we compare with our DNS data. 

The fourth parameter  in Eq.~(\ref{eq:muc}) is the scale $\Vc$. It depends upon the difference of the two Stokes numbers.
We follow Ref.~\cite{meibohm2017relative} and write
\begin{equation}
\label{eq:defstmtheta}
\quad \theta=\frac{ |\Sto-\Stt|}{\Sto+\Stt}\,.
\end{equation}
The white-noise model predicts that \cite{meibohm2017relative}
\begin{equation}
\label{eq:rcvctheta}
\Vc\! \propto\! \theta
\end{equation}
at small $\theta$.
In this case, the power-law tails of the distribution (\ref{eq:muc})  are expected to contribute to the relative velocity moments. According Eq.~(\ref{eq:muc}), the tails of the distribution beyond $\Vc$ are simply those of the mono-disperse system. 

Eq.~(\ref{eq:muc}) implies that the distribution of separations becomes uniform in $R$ for 
$R< \Rc$, as predicted in Refs.~\cite{Bec05,Chu05}. 
Their spatial scale $\Rc$ is thus related to our velocity scale as follows:
\begin{equation}
\Rc \equiv \Vc/z^\ast\,,
\end{equation}
and therefore $\Rc\propto\theta$ at small $\theta$.

\subsection{Moments of relative velocities}
\label{sec:moments_theory}
Theoretical predictions for $\langle |\VR|^p\rangle$ are obtained by integrating the distribution $\mP$, as determined by Eq.~(\ref{eq:mp}). We first quote the results when $\theta$ is small, when the distribution exhibits
a clear power law. This power law is cut off at small relative velocities at $\mbox{max}(\Vc,\zast R)=z^\ast\mbox{max}(\Rc,R)$, consequently
the result for $\langle |\VR|^p\rangle$ depends on whether $R>\Rc$ or not.
When  $R>\Rc$ we find
\begin{equation}
\label{eq:mp1}
m_p(R)=b_p R^{\muc+p-1}+c_p R^{d-1}\,,
\end{equation}
with
\begin{align}
\label{eq:bpcp1}
b_p &= -\frac{\mathscr{N}(1\! + \!d \!- \muc){z^*}^{p+1}}{(p\!+\!1)(\muc\!-\!d\!+\!p)}\,,\\
c_p&=\frac{\mathscr{N}{z^*}^{p+1}{(\tfrac{V_0}{\zast})}^{\muc -d + p}}{\muc-d + p}\,,
\nonumber
\end{align}
where $\mathscr{N}$ is the normalization factor in Eq.~(\ref{eq:muc}).
For large  values of $p$, the coefficients     
$b_p$ 
and  $c_p$ are sensitive to  the form of the distribution beyond the cutoff $\zast$, which depends on the nature of the turbulent fluctuations. Also, the value of $\muc = D_2(\Stm)$ is not universal, and neither is the parameter
$z^\ast$. 
The second term in \Eq{eq:mp1} appears due to presence of 
singularities (of the gradient of particle velocity) 
called caustics~\cite{wilkinson2005caustics,wilkinson2006caustic} for non-zero values of $\St$. 
In other words, the presence of caustics imply that while the distance between two nearby particles 
goes to zero their relative velocities can remain order unity. 
Whereas, in the absence of caustics, the particle velocity field remains smooth -- relative 
velocity of two particles  goes to zero as separation between them goes to zero, 
this gives rise to first term in \Eq{eq:mp1} (see Ref.~\cite{gustavsson2014relative} for more discussion).

The $R$-dependence
predicted by Eq.~(\ref{eq:mp1}) is universal. It is equal to the scaling form of $m_p(R)$ for  identical particles \cite{Gus12}, as expected for small $\theta$. But for particles with different Stokes numbers the coefficients
$b_p$ and $c_p$ depend upon $\theta$, although only through the global normalization constant $\mathscr{N}$. 
The scale $\Vc$ does not enter explicitly because $R>\Rc$.

Now consider  $R < \Rc$. Then
 the uniform part in Eq.~(\ref{eq:muc}) dominates the moments.
At $R<\Rc$, particles of two different sizes $a_1$ and $a_2$ move approximately independently from each other. In this case the moments take the form:
 \begin{align}
 \label{eq:moments_diff}
m_p(R)
& \sim c'_pR^{d-1}\,,
 \end{align}
 with
 \begin{align}
 \label{eq:cpp}
c'_p=c_p-\frac{{\mathscr{N}}(1 + d - \muc)(\Vc/z^\ast)^{\muc -d + p} {z^*}^{p+1}}{(\muc - d + p) (p+1)}\,.
\end{align}
For $p=1,2,3,\ldots$ one finds that $c_p' < c_p$  for heavy particles in incompressible turbulence at not too large Stokes numbers [DNS show that $D_2>d-1$, and 
that $D_2< d+1$ for not too large Stokes  numbers, see Eq. (\ref{eq:muc2})].
The moments for larger $\theta$ are nevertheless usually larger than those for $\theta\to 0$, because
the term  $b_pR^{D_2+p-1}$ makes a large negative contribution unless $R$ is extremely small, and
this term is absent in Eq.~(\ref{eq:moments_diff}).  In general, if $\Stm$ is small enough so that caustics are rare, then Eq.~(\ref{eq:moments_diff}) can give a contribution for different particles
that is much larger than for identical particles,
leading to a significantly higher collision rate.
The dependence on $R$ is of the same form as the caustic contribution in Eq.~(\ref{eq:mp}) in the limit
$\theta\to 0$. 

Finally consider larger values of $\theta$, large enough so that the power laws in Eq.~(\ref{eq:muc}) disappear.
In a Gaussian white-noise model the distribution $\mP(R,\VR)$ is Gaussian 
in this limit \cite{meibohm2017relative}. 

\begin{figure*}
\begin{center}
\includegraphics[width=0.45\linewidth]{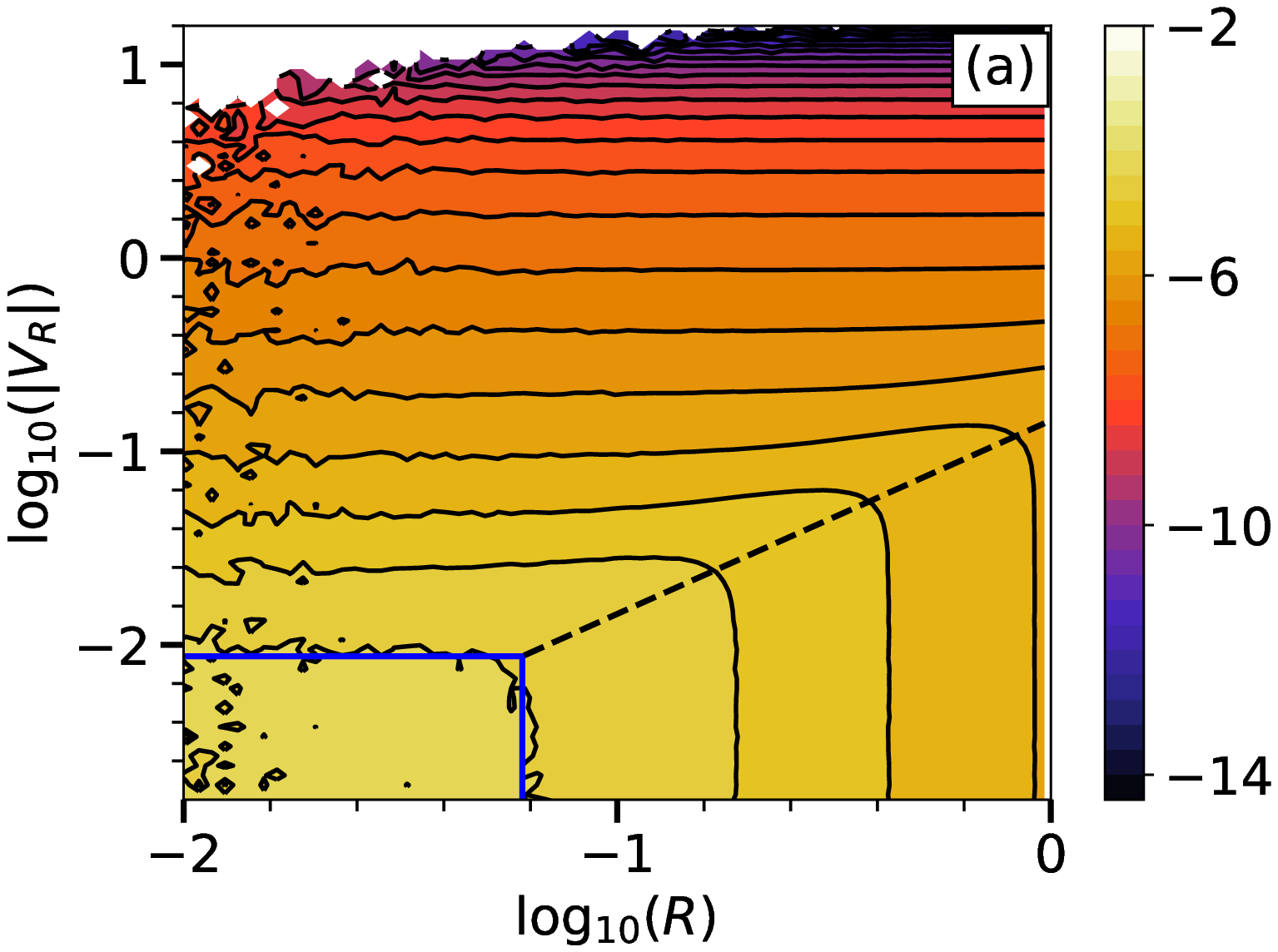}
\includegraphics[width=0.45\linewidth]{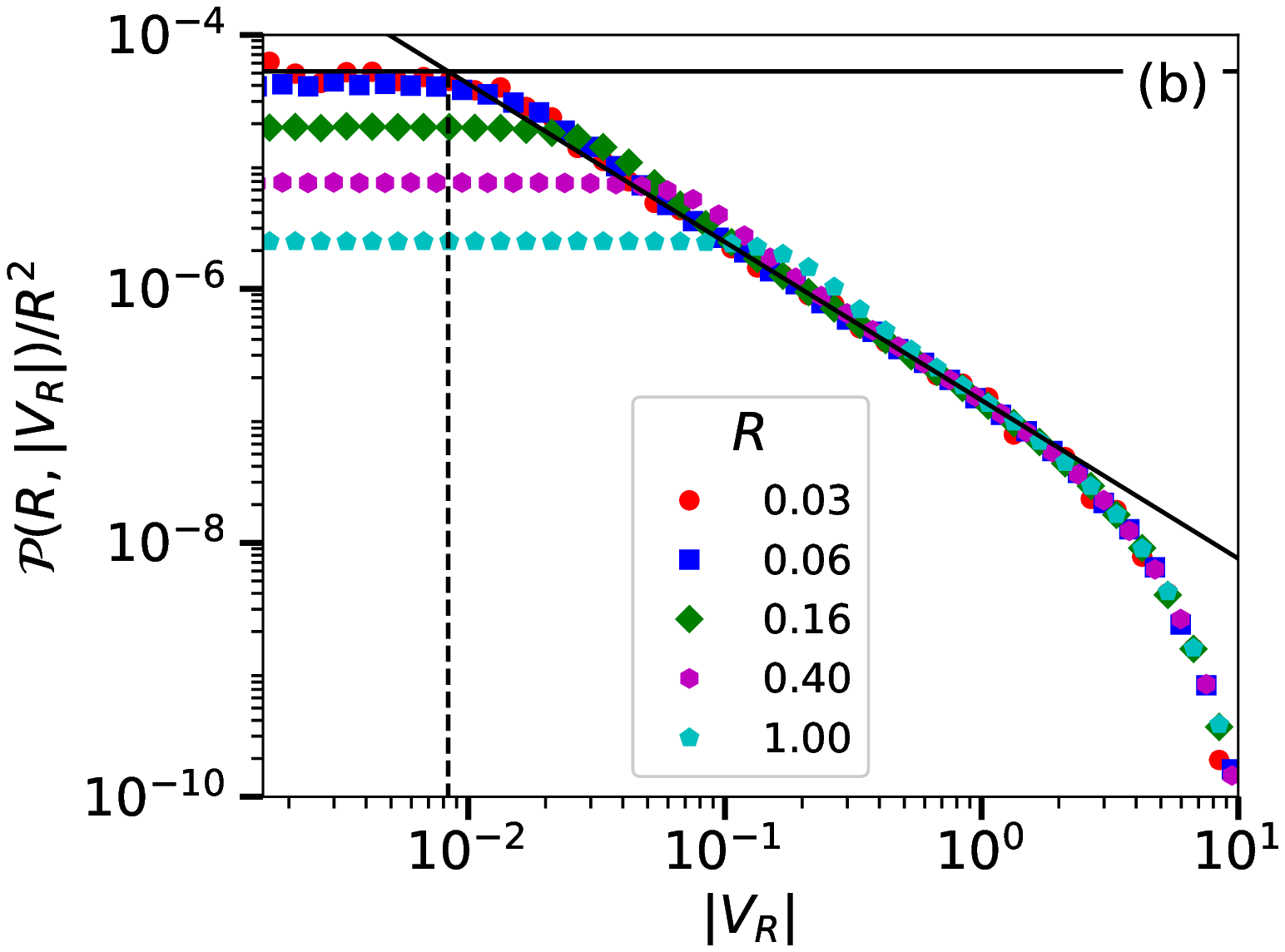}\\
\includegraphics[width=0.45\linewidth]{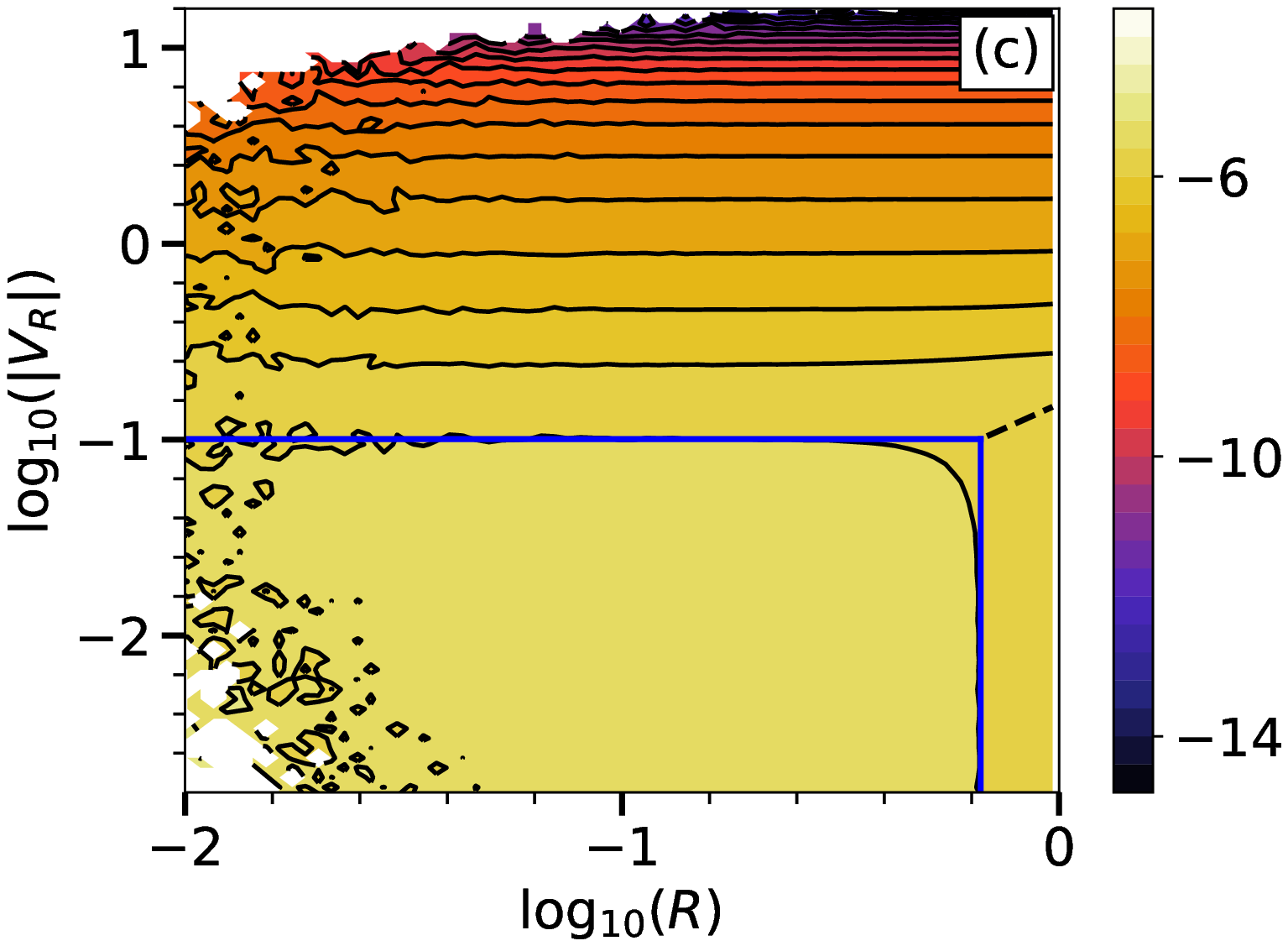}
\includegraphics[width=0.45\linewidth]{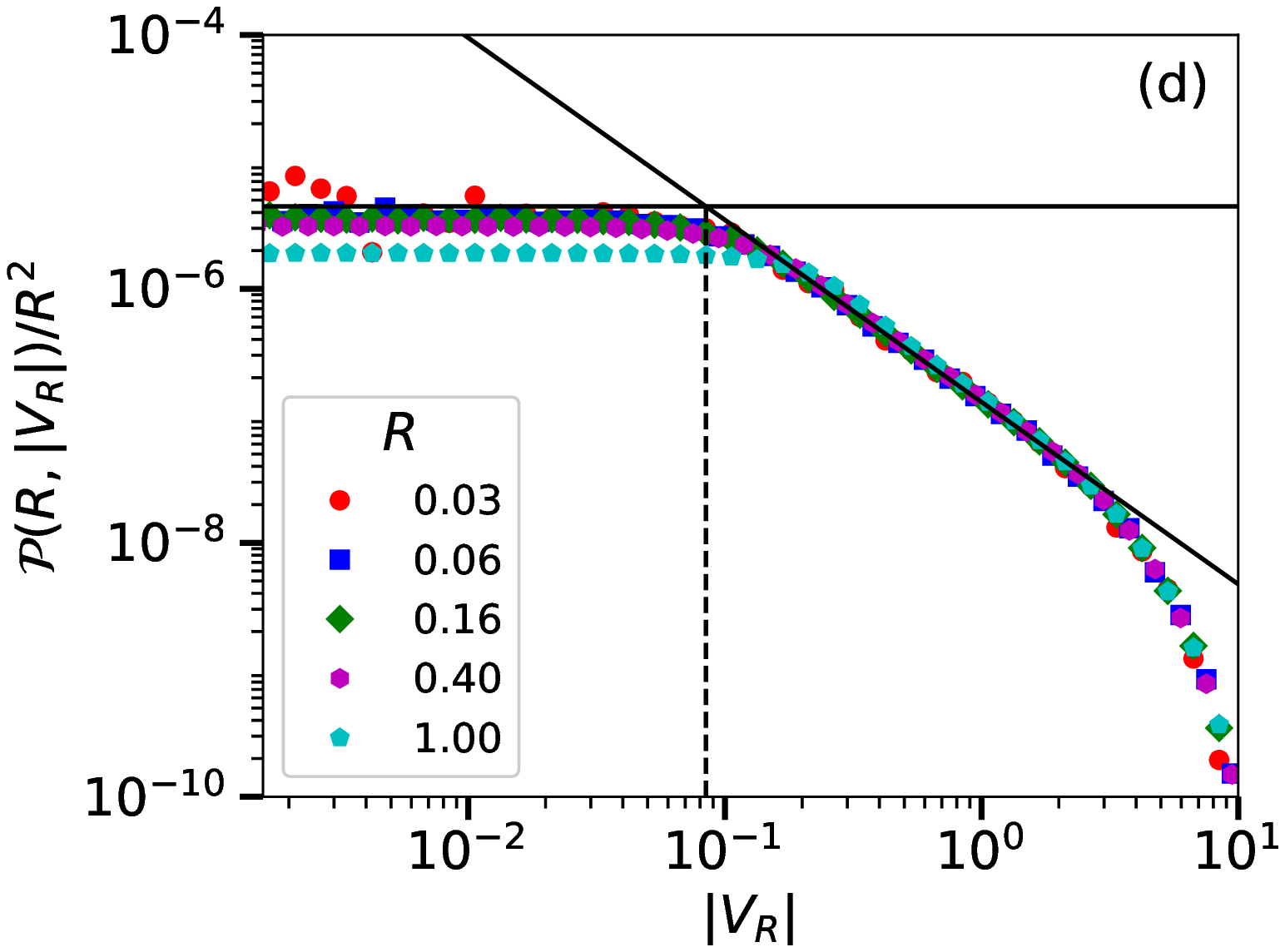}\\
\includegraphics[width=0.45\linewidth]{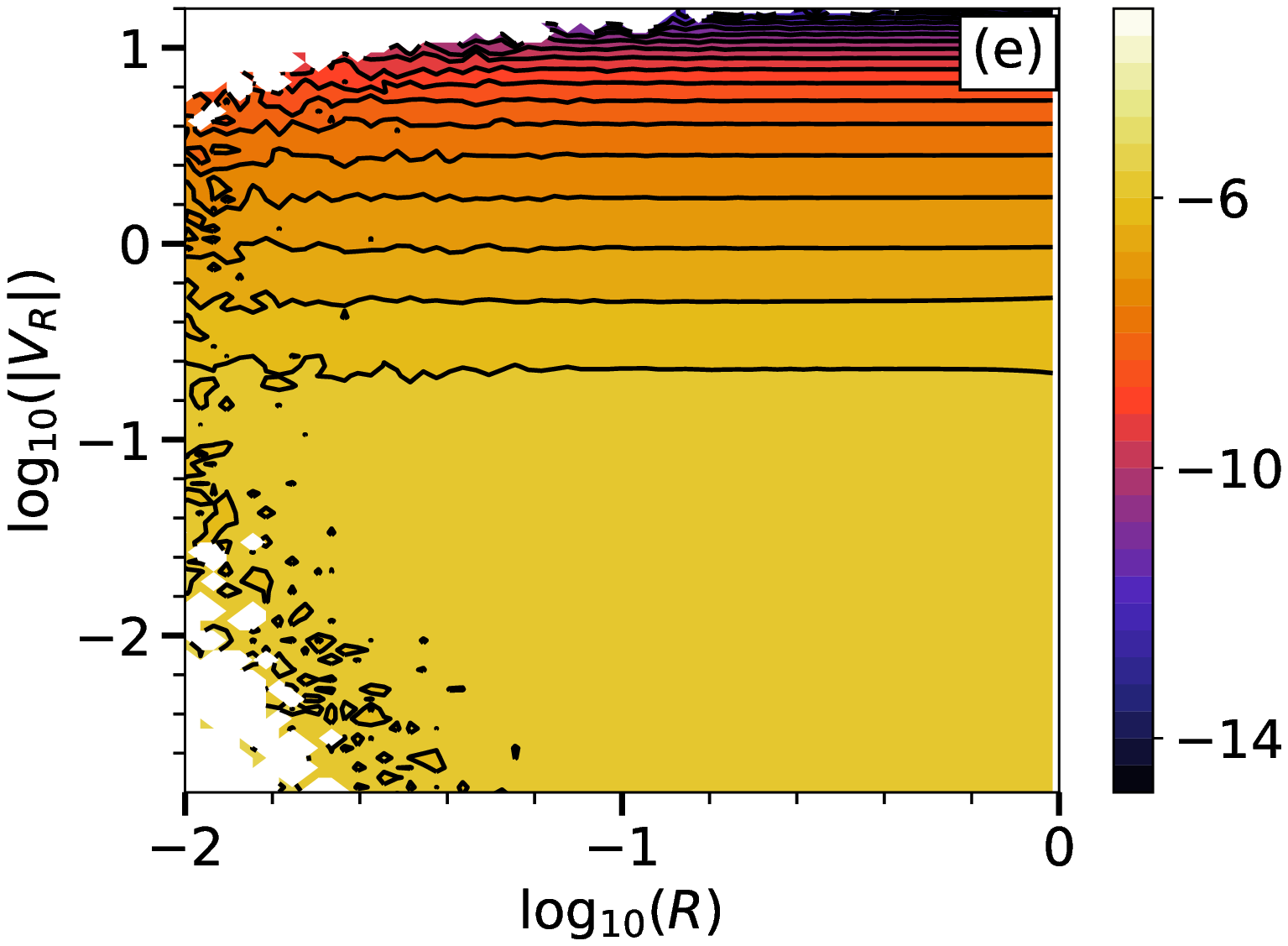}
\includegraphics[width=0.45\linewidth]{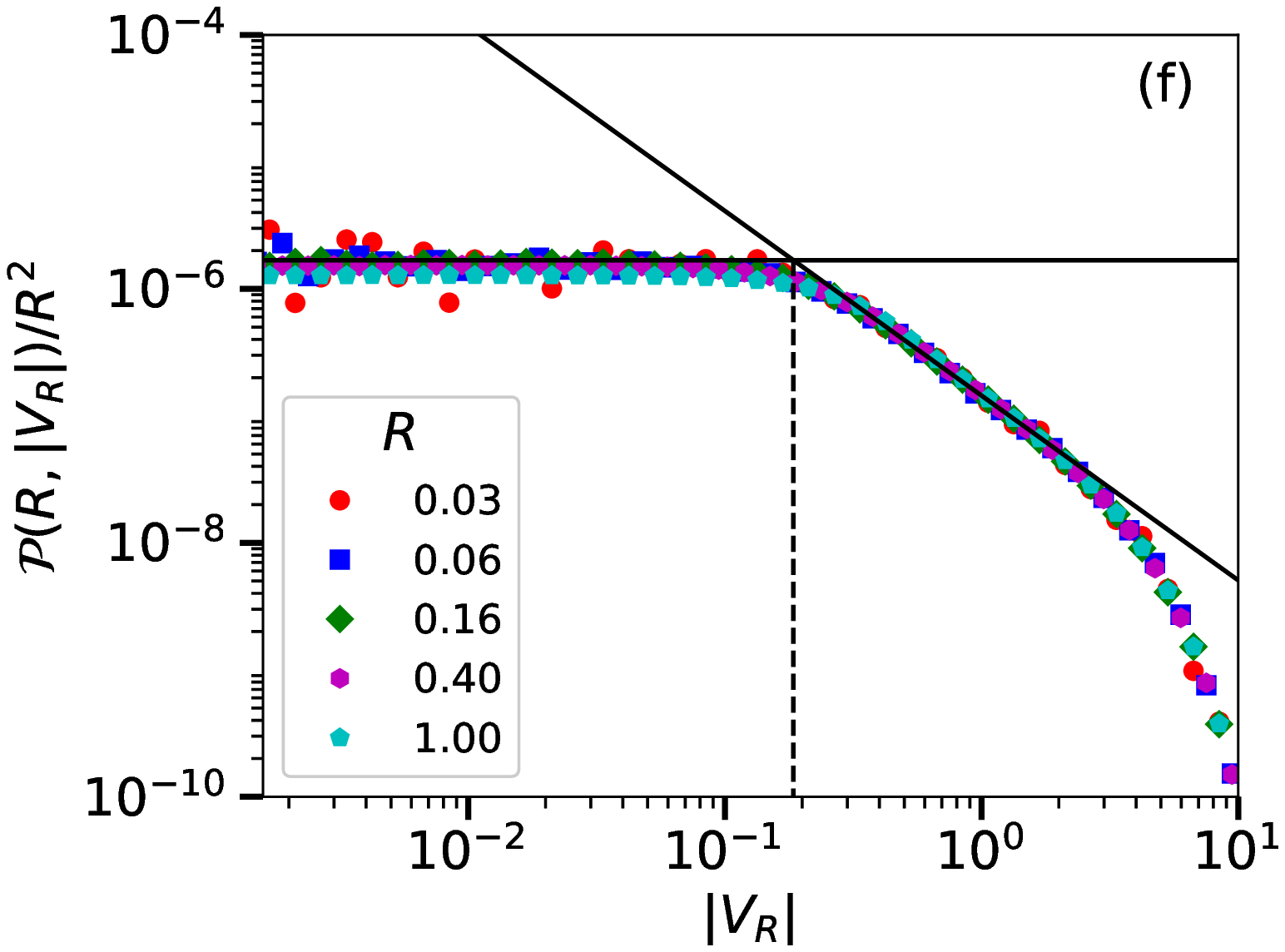}
\caption{(color online)  DNS results for joint distribution $\mP(R,\VRa)$ of $R$ and $\VRa$, divided by $R^2$. 
Parameters:  $\Stm=2$ and $\theta=0.005$ (top row),
  $\theta=0.05$ (second row), and $\theta=0.1$ (bottom row). First column: Contour plots
of $\mP(R,\VRa)/R^2$ 
   color coded according to $\log_{10}[\mP(R,\VRa)/R^2]$.  The blue lines in
  the bottom left corner of these plots show the scales $\Rc$
  and $\Vc$ (see text). The dashed lines show the theoretical matching condition $\VRa = z^\ast R$ (see text). 
  Second column: plots of $\mP(R,\VRa)/R^2$ as
  functions of $\VRa$ for different values of $R$ as indicated in the panels. 
  Also shown  are fits (solid lines) to the theoretical power-law
  prediction $\VRa^{\muc-4}$, Eq.~(\ref{eq:muc}),  to 
  determine $\muc$ as a function of $\Stm$. 
 The crossover between the  approximately uniform (broad Gaussian) part
    at small $\VRa$ (and small $R=0.03,0.06$, horizontal solid lines) and the power-law
 at intermediate $R$ sets the scale $\Vc$ (dashed vertical lines).
}
\label{fig:jpdf}
\end{center}
\end{figure*}

\subsubsection{Very dissimilar pair of particles}
When one of the particles has a very small Stokes number, $\St_2\ll1 $ say, we can evaluate
the coefficient $c'_p$  term in  Eq. (\ref{eq:moments_diff})
 in terms of single-particle observables. We  now outline the calculation for $p=2$.
 When
 $\Stt \ll 1$, we can expand the equation of motion up to leading order in
$\Stt$ to obtain the velocity of the second particle:
\begin{eqnarray}
\vvt &\approx& \uu(\xx,t) - \ma A\cdot\RR - \Stt {\DDt{\uu}(\xx+\RR,t)}\,.
\end{eqnarray}
The relative velocity between two particles can then be written as
\begin{eqnarray}
\VV(\RR)  &\approx &  \vv - \uu(\xx,t) \nonumber \\
                && + \ma A\cdot\RR  + \Stt {\DDt{\uu}(\xx+\RR,t)}  \/.
\label{eq:v1}
\end{eqnarray}
The first line of the right-hand-side of \Eq{eq:v1} is $\Sto$ times the
acceleration of a {\it single} particle; at small $|\RR |$ and
$\Stt$ this is the leading order contribution to the relative
velocity.
The distribution of the acceleration has been studied
extensively and is known to have exponential tails~\cite{bec2006acceleration,aks_thesis}.
This information allows us to approximately relate  the structure functions to single-particle averages, as shown below.

We assume that to calculate $\langle\VR^2\rangle$ for $R$ much smaller than $\Rc$ 
it is sufficient to consider one component of $\VV$. 
Consider one component of \Eq{eq:v1}, square both sides of the resultant equation and then 
take steady-state averages.
Assuming that  $R\ll 1$ we obtain:
\begin{align}
\nonumber\bra{\VR^2} &\approx {\frac{1}{3}}\left[{\bra{\uu^2}} - {\bra{\vv^2}}\right]\Big( 1-
 2 \frac{\Stt}{\Sto} \Big)\\
&{-{\frac{2}{3}}\Stt\langle(\uu-\vv)\cdot\ma A\cdot(\uu-\vv)\rangle}\,.
\label{eq:vrms}
\end{align}
All averages on the r.h.s. of Eq.~(\ref{eq:vrms}) are evaluated for a single
particle with Stokes number  $\Sto$. The only $\St_2$-dependence appears in the
prefactors on the r.h.s. of Eq.~(\ref{eq:vrms}). We note that there is no $R$-dependence
(since all averages are single-particle averages). This is the result of neglecting 
the gradient term  $\ma A\cdot\RR$ in the equation for the particle separations. As explained in Section
II.A of Ref.~\cite{meibohm2017relative} this is allowed provided that $R<\Rc$. 
But note that in Ref.~\cite{meibohm2017relative} the {\em white-noise model} 
was analyzed, while Eq.~(\ref{eq:vrms}) applies to a turbulent flow. 

\section{ DNS results}
\label{results}

\subsection{Distribution of relative velocities and separations}
\label{sec:distribution_dns}
Fig.~\ref{fig:jpdf} shows
a comparison between  the theory Eq.~(\ref{eq:muc}) and our DNS results for $\mP(R,\VR)/R^{2}$ for different values of 
$\theta$. 
The first column  of panels in this Figure shows contour plots of $\mP(R,\VR)/R^{2}$.
As predicted by the theory (\ref{eq:muc}), there is a region in the $R$-$\VR$ plane where the distribution is a broad Gaussian. In a log-log plot this appears as an approximately uniform region where $\mP/{R^{2}}$ is approximately constant. Outside this region, and  for small values of $\theta$, the equidistant contour lines show that the distribution exhibits  the power laws, as predicted by the theory.

To analyze the power laws in relative velocities in more detail, the  second column 
of panels in Fig.~\ref{fig:jpdf}  shows plots of $\mP(R,\VR)/R^2$ as
functions of $\VRa$ for several different values of $R$. We can clearly distinguish the power-law 
from the broad Gaussian at small $|\VR|$, where $\mP/{R^{2}}\approx$ const. 
Eq.~(\ref{eq:muc}) says that the cross over between these two behaviors occurs at $\mbox{min}(\Vc,\zast R)$. We estimate this cross-over velocity scale by drawing two lines: a horizontal one at small $\VRa$,
and a power-law fit for larger $\VRa$.  The scale at which these two lines intersect is our estimate of the
cross-over scale. For small values of $R$ the fits yield a velocity scale that is independent of $R$,
this is $\Vc$. For slightly larger values of $R$, the velocity scale is proportional to $R$, as predicted by theory,
and  the constant of proportionality defines the parameter $\zast$.

Dissipation-range theory  \cite{meibohm2017relative} says that $\Vc =c\, \theta$ for small $\theta$, but the theory does not determine
the constant of proportionality $c$. This constant is system specific, as is the value of $z^\ast$. In the white-noise limit these parameters can be calculated analytically \cite{gustavsson2014relative,meibohm2017relative}, but not in general. 

\begin{figure*}
\begin{center}
\includegraphics[width=0.32\linewidth]{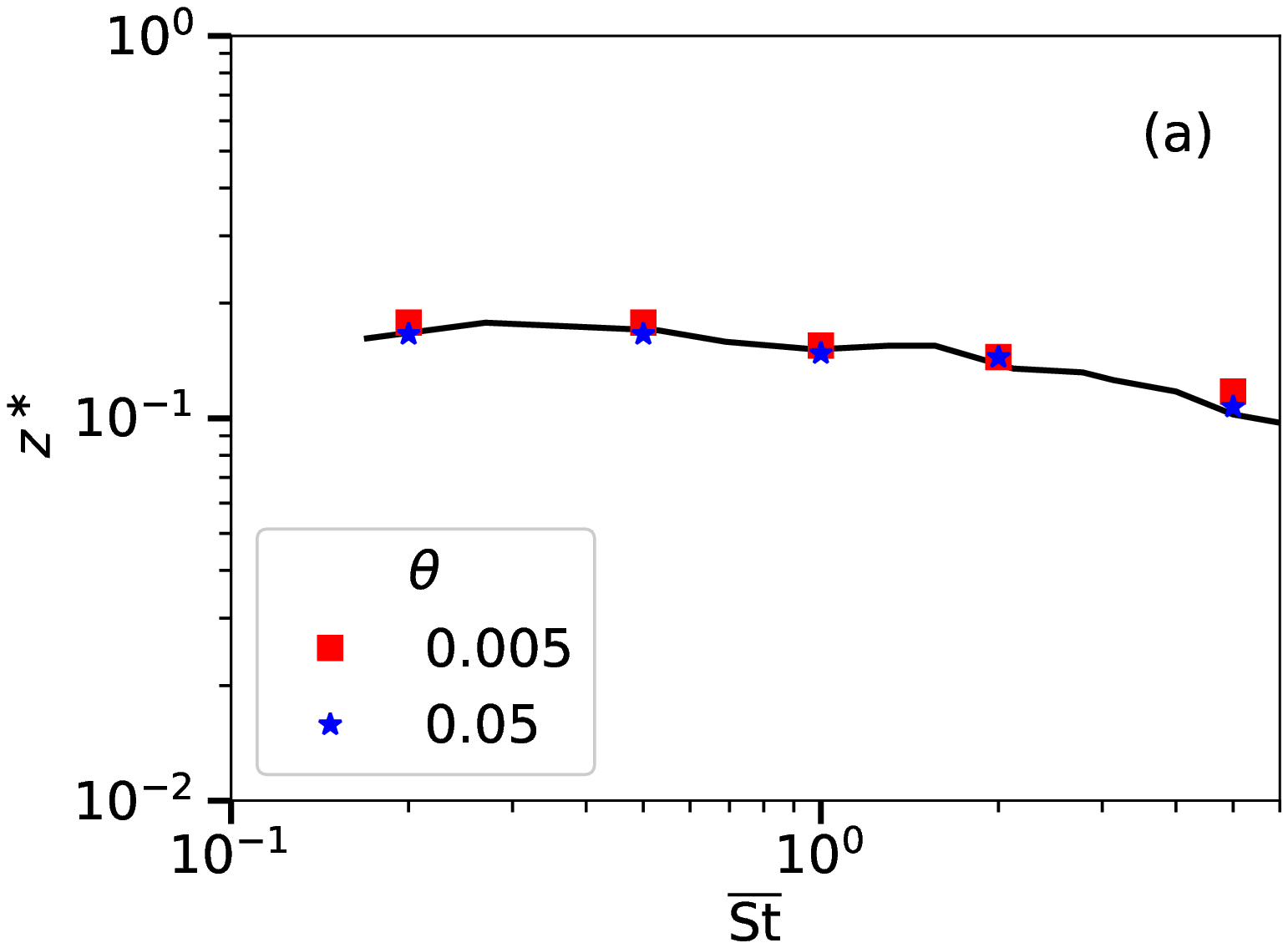}
\includegraphics[width=0.32\linewidth]{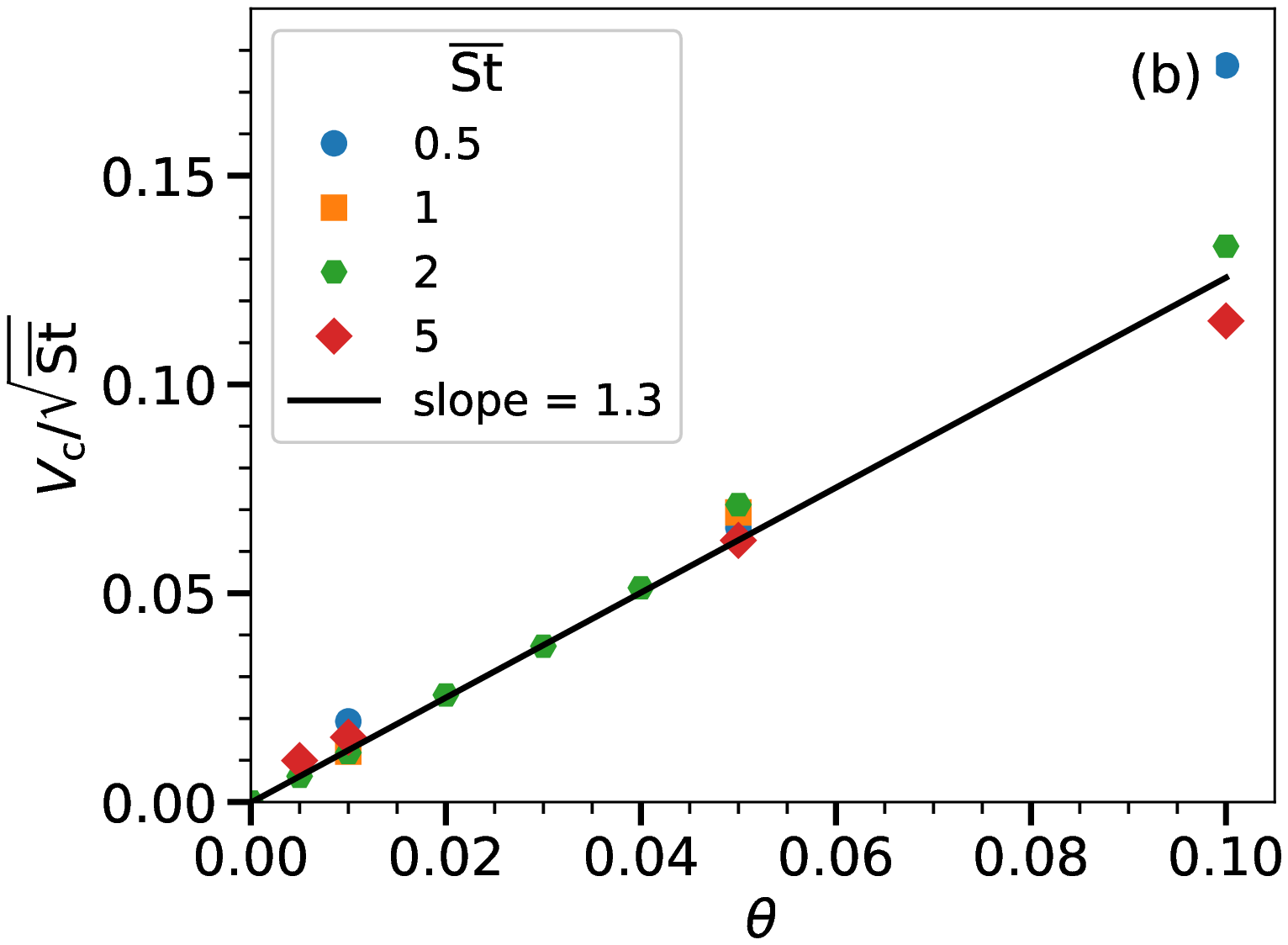}
\includegraphics[width=0.32\linewidth]{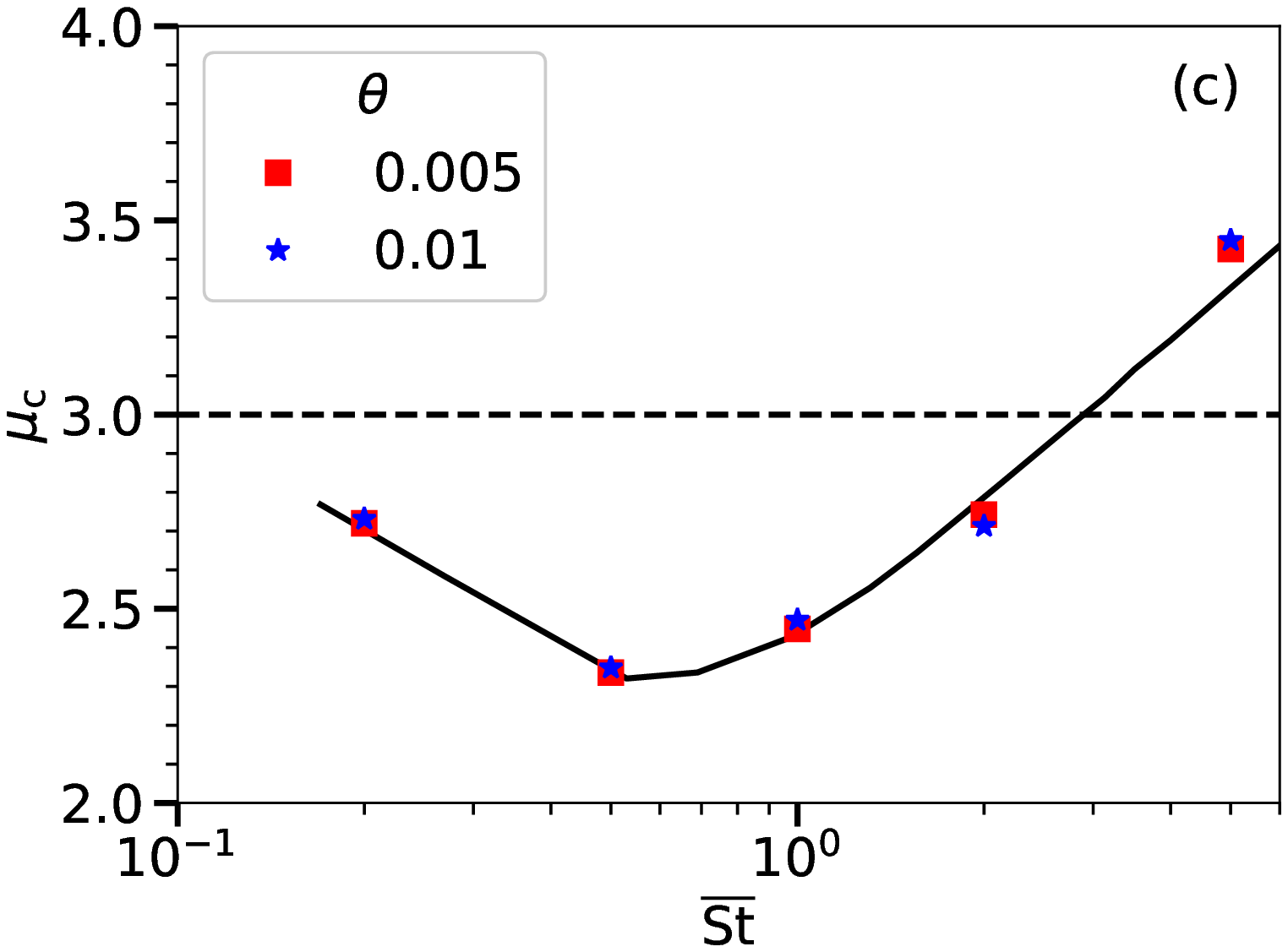}
\caption{
(color online) Estimates of the parameters $z^\ast$, $\Vc$, and $\muc$, obtained
 from the DNS results for $\mP(R,\VR)$  shown in Fig.~\ref{fig:jpdf}.
(a) Scale $\zast$ as a function of $\Stm$, for two different
values of $\theta$ (symbols). The solid black line  is the  estimate
for identical particles, taken from the DNS of Ref.~\cite{bhatnagar2018statistics}. 
(b) Scale $\Vc$ as a function of $\theta$ (symbols), for different values of $\Stm$.
The solid black line shows a linear dependence upon $\theta$ with fitted prefactor $1.3$.
(c) Exponent $\muc$ as a function of $\Stm$ for two different values of $\theta$ obtained by
  power-law fits to DNS results for $\mP(R,\VR)$ at fixed $R$, see Fig.~\ref{fig:jpdf}. The solid black line is the phase-space
  correlation dimension $\Dtwo$ of the fractal attractor for identical particles with Stokes number $\Stm$, taken from Ref.~\cite{bhatnagar2018statistics}. 
  }
\label{fig:params}
\end{center}
\end{figure*}

Therefore it is important to determine these constants by DNS.
The results are shown in Fig.~\ref{fig:params}. Panel (a)  shows that $\zast$ is essentially independent of $\theta$, while
panel (b) demonstrates that $\Vc$ is proportional to $\theta$ at small $\theta$, as predicted by the theory.
 Fig.~\ref{fig:params}(b) also shows that the prefactor depends on $\Stm$ as $\Stm^{1/2}$, at least
 for the parameters simulated. This follows from the fact that the DNS data for  $\Vc\,\Stm^{-1/2}$
 collapse onto a single line. 
However,  there is no theoretical explanation for this result, as far as we know.

Fig.~\ref{fig:params}(c) shows the power-law exponents $\muc$.
We  extracted $\muc$ for different values of $\Stm$ and
for two different values of $\theta$ by fitting power laws to the DNS results for the
distribution of relative velocities.
Panel (c) shows the resulting exponents $\muc$ together with $\Dtwo$ for the case $\Sto=\Stt$ from Ref.~\cite{bhatnagar2018statistics}.
Up to the numerical accuracy in our DNS we find  for $D_2<4$ that $\muc=D_2$, independent of $\theta$ for small values of $\theta$.
The phase-space correlation dimension $\Dtwo$ has a characteristic minimum at $\Stm$ of order unity and monotonously approaches the spatial dimension $d$ for small $\Stm$ and the dimensionality of phase space, $2d$, for large $\Stm$ [see Fig.~\ref{fig:params}(c)].

In summary we observe good agreement between our DNS and the theory, Eq.~(\ref{eq:muc}), in particular
for small $\theta$. As $\theta$ increases, the velocity scale $\Vc$ grows so that the range
of the power law between $\Vc$ and $V_0$ becomes smaller. For large enough values of $\theta$,
the power laws disappear. In this limit the distribution is a broad Gaussian, approximately uniform. 
In our log-log plots, $\mP/{R^{2}}$ is approximately constant in this region.

\subsection{Moments of relative velocities}
\label{R:mean}
Fig.~\ref{fig:moments} summarizes our DNS results for the moments of relative 
velocities as a function of particle separation. 
Panel (a) shows DNS results for $m_0(R)/R^2$ as a function of $R$ (symbols), while  panel (b) shows $m_2(R)/R^2$,
also as a function of $R$.
The parameters are given in the Figure caption.  Also shown
is the scaling of the smooth contribution predicted by \Eq{eq:mp1} (solid line). 
Dashed vertical lines correspond to the scale $\Rc = \Vc/\zast$. 
The parameters $\Vc$, $\muc$, and $z^\ast$
were determined separately, as described in Section \ref{sec:distribution_dns}. 

As predicted by \Eq{eq:mp1},  the moments scales as $R^{d-1}$ for $R < \Rc$.
For $R>\Rc$ smooth contribution dominates for $m_0(R)$ for both values of $\Stm$,
whereas for higher order moment $m_2(R)$ smooth contribution dominates only
for the smaller mean Stokes number. For larger  mean Stokes number, the caustic contribution $c_p R^{d-1}$
swamps the smooth part for $R$ below $\Rc$. 
In limit the relative-velocity moments $m_p(R)$ are dominated
by the singular $R^{d-1}$-contribution provided that $p$ is large enough. While the $R$-dependence of this contribution
is the same for identical particles and for particles with different
Stokes numbers, the physical origin of this  power law is slightly
different in the two cases.
For identical particles, the singular term is caused by caustics \cite{falkovich2002acceleration,wilkinson2005caustics,wilkinson2006caustic}. For particles with different Stokes numbers, by contrast,  the singular contribution is due to the uncorrelated motion between nearby ($R <\Rc$)
particles with different Stokes numbers \cite{meibohm2017relative}. 

\subsubsection{Very dissimilar pair of particles}
\begin{figure*}
\begin{center}
\includegraphics[width=0.32\linewidth]{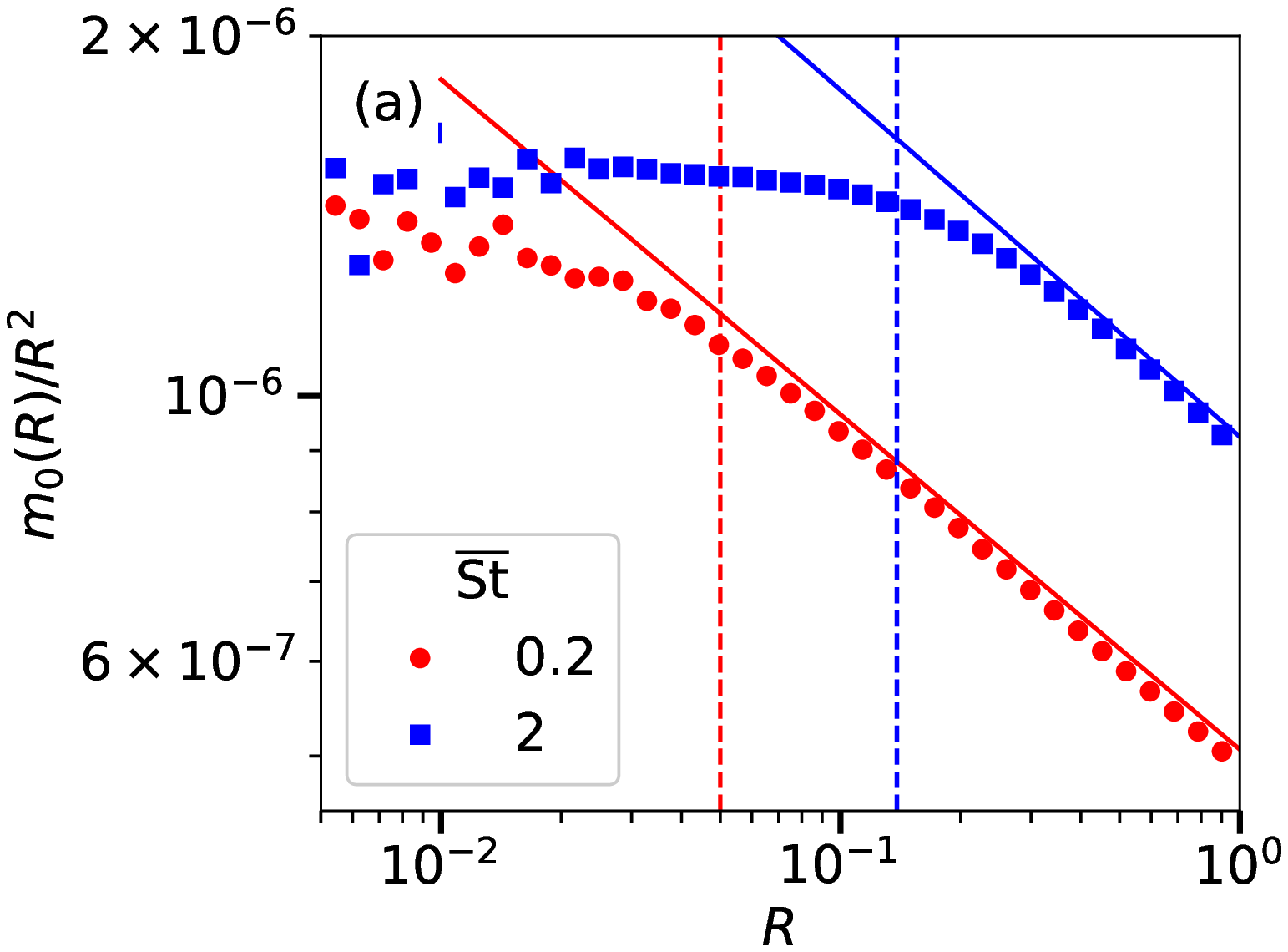}
\includegraphics[width=0.32\linewidth]{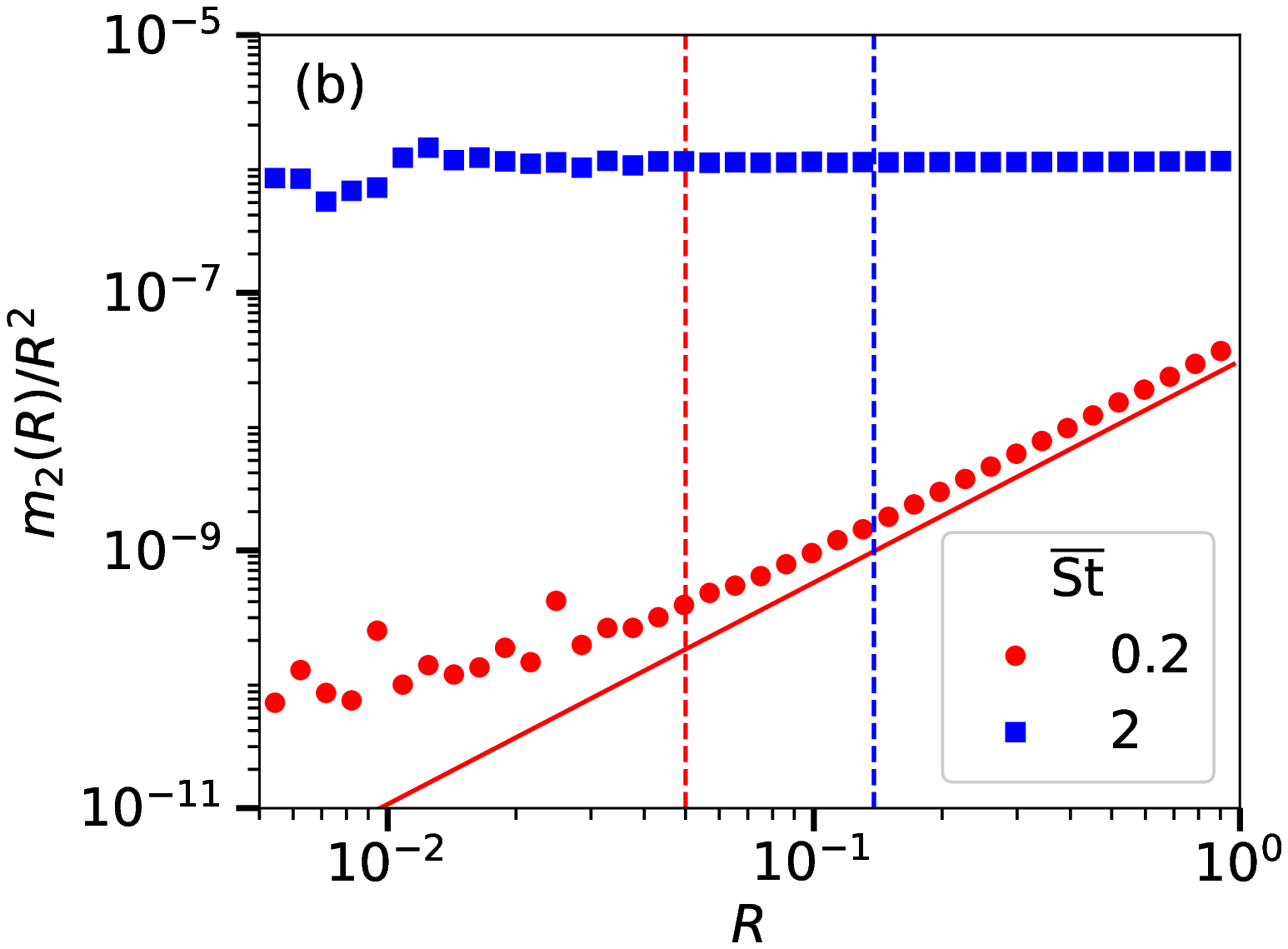}
\includegraphics[width=0.32\linewidth]{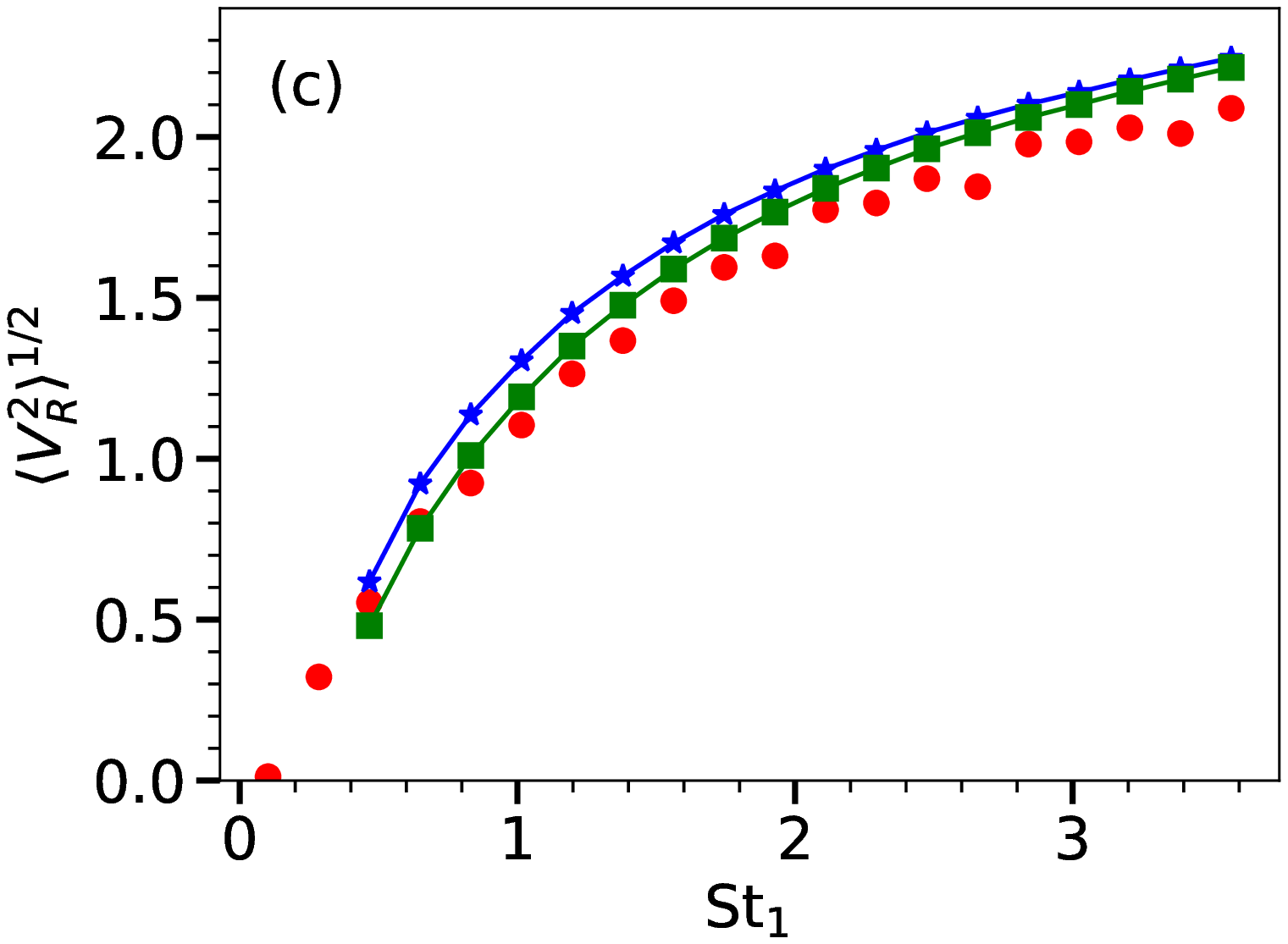}
\caption{(color online) DNS results for moments of relative velocities as a function
of particle separation $R$.  (a) Zero-th moment $m_0(R)$ and (b) second moment divided by $R^2$,
for $\Stm=0.2$ and $2$,  and $\theta=0.01$ (symbols). Solid line shows the scaling of
smooth contribution in \Eq{eq:mp1}. The scale $\Rc=\Vc/\zast$ is
indicated by the dashed vertical line.
(c) Root-mean-square radial velocity  $\langle \VR^2\rangle^{1/2}$ 
for $R = \ao+\at$ plotted as a function of $\Sto$ for $\Stt=0.1$ (red
circles). 
The first term of theoretical estimate, \Eq{eq:vrms}, is plotted as blue $\star$
(joined with a blue solid line). The full expression \Eq{eq:vrms}, is plotted with green $\blacksquare$
(joined by a green solid line).
}
\label{fig:moments}
\end{center}
\end{figure*}

Fig.~\ref{fig:moments}(c) shows DNS results for $\langle \VR^2\rangle$ at the collision radius
$R=a_1+a_2$ for $\St_2\ll1$ as a function of $\St_1$ (red circles),
that is for large values of $\theta$. Also shown is the theoretical expression,
\Eq{eq:vrms} (green squares). The averages on the r.h.s. of \Eq{eq:vrms}  are determined
by DNS, by averaging along heavy-particle paths in the steady state.
The agreement  is good at small values of $\St_1$, but we observe 
deviations at larger values of the Stokes number. 
It is possible that this is due to higher-$\St_2$-terms neglected
in (\ref{eq:vrms}). Plotting only the first term of \Eq{eq:vrms} yields slightly different results, although the deviations are smaller than those between the full theory and the DNS results. 

We have checked that the gradient term $\ma A\cdot \RR$ in the equation of motion for the separation $\RR$
is negligible. For all data points shown, $\theta$ is large enough so that  $a_1+a_2$ is much less than $\Rc$.
In this range the DNS results do not depend upon $R$. This is the plateau region seen in Fig.~\ref{fig:moments}(a).

\section{Discussion}
Our results show in agreement with the theory that the distribution of relative velocities
is non-Gaussian when $\theta$ is small. For a fairly wide range
of $\theta$ (up to $\theta \sim 0.1$), the distribution has power-law tails $\sim|\VR|^{\muc-4}$ at small separations. The dissipation-range theory predicts that the  exponent $\muc$ is determined 
by the {\em phase-space} correlation dimension $D_2(\Stm)$ for a mono-disperse system
with Stokes number $\Stm$ [Eq.~(\ref{eq:muc2})]. In our simulations, the numerical values of $\muc$ vary  from approximately $2.4$ to $3.5$, and in this range there is good agreement between the theory and the 
numerical values of $\muc$ obtained from the DNS~\footnote{This agreement should be understood 
in the following manner. The theory does not allow a calculation of $\muc$ from first principle,
but it shows that $\muc = D_2(\Stm)$ for small $\theta$. This is indeed what we confirm from DNS}.

In the astrophysical literature, several papers have reported DNS results for the distribution of relative particle velocities 
\cite{pan2013turbulence,pan2014distri,ishihara18}. These authors attempted to fit the distributions to stretched exponentials,
of the form $\exp[-(|\VR|/\beta)^\gamma]$ with fitting parameters $\beta$ and $\gamma$. The parameter
$\gamma$ is usually quoted to be smaller than unity. This law is neither consistent with our power-law
predictions, nor with the large-$\St$ prediction from Ref.~\cite{Gus08b}. We have reanalyzed the data
in Fig.~12 of Ref.~\cite{ishihara18} for the two smallest Stokes numbers, and find clear power laws over one decade
of $\VR/u_\eta$, with exponents $\mu_c-4$ in good agreement with the dissipation-range theory (the values
of $\muc$ were obtained from the plots of the pair correlation function in Fig.~8 of the same paper).

We remark that the distribution of relative velocities in bidisperse suspensions was recently studied 
in Ref.~\cite{dhariwal2018small}. This study did not report power-laws for the distribution of relative velocities.
As our results show, possible reasons for the absence of power laws are, firstly, that the distributions
were calculated at quite large separations (of the order of the Kolmogorov length, $R\sim\eta$).
Secondly, the values of $\theta$ were quite large, too large to see power laws as our theory and DNS data
demonstrate.

Pan and Padoan \cite{pan2013turbulence} did not plot the radial relative velocity $\VR$ (that determines
how particles approach each other), but instead the RMS relative velocity $V_{\rm rms}\equiv\sqrt{V_1^2+V_2^2+V_3^2}$. 
The power law of the distribution of $V_{\rm rms}$  has a different exponent~\cite{gustavsson2011distribution,gustavsson2014relative}:
$|V_{\rm rms}|^{\muc-2d}$. We have compared this prediction with the data shown 
in Fig.~14 of Ref.~\cite{pan2013turbulence}. There is a clear power law, with exponent 
$\approx -3.7$ for $\St =1.55$. Theory says that the exponent should equal $D_2-6$, but 
Ref.~\cite{pan2013turbulence} does not give values for the fractal correlation dimension $D_2$.
Estimating $D_2$ from our data at $\St=1.55$ (albeit at a different Reynolds number),
we find $D_2-6\approx -3.4$, in reasonable but not perfect agreement with the DNS results
of Ref.~\cite{pan2013turbulence}.

Ishihara {\em et al.} state that their distribution approaches a Gaussian when $\theta$ is not small.
This is consistent with theory \cite{meibohm2017relative}, predicting a broad Gaussian for the body of the distribution. 
In our log-log plots, \Fig{fig:jpdf}, the broad Gaussian appears as a region where
 $\mP/{R^{2}}$ is approximately constant. When $\theta$ is large enough, this region extends
 out to $V_0$,  approximately equal to the RMS turbulent velocity, $\urms$. The form of the far tails beyond $V_0$
 is difficult to determine, because the tails describe rare events, and since there is no theoretical prediction
 apart from the law predicted in Ref.~\cite{Gus08b}. Yet this applies only at large Stokes numbers, and when
 there is a well-developed inertial range. 
 
 In both cases, when $\theta$ is small and when it is large, the RMS relative velocity is determined by 
 the upper cutoff, $V_0$. We have simply set $V_0 = \urms$ here, but this is a simplification. 
In general, the upper cutoff $V_0$ must also depend on particle inertia (Stokes number).
We have neglected this dependence here. Taking $V_0=\urms$ implies that 
the moments of particle relative velocities depend on the Reynolds number
  $\Rey$ when determined by the upper cutoff $V_0$, since 
$\urms/\ueta \propto \Rey^{1/4}$~\footnote{This can be derived using the Kolmogorov scaling
$\urms/\ueta \sim (l_f/\eta)^{1/3}$, where $l_f = 1/\kf$ is forcing scale.}.  
  With our present computational capabilities
we cannot explore such a weak dependence on $\Rey$; hence we
have concentrated our efforts on a single value of $\Rey$.
Experimental data~\cite{dou+bra+ham+etal18} confirms that the $\Rey$-dependence is quite weak.

Ishihara {\em et al.} \cite{ishihara18}, on the other hand, computed RMS relative particle velocities for different values of $\Rey$ 
(Fig.~3 in their paper), obtaining a fairly strong dependence on $\Rey$. A possible explanation of this result is
that Ishihara {\em et al.} evaluated $\langle \VR^2\rangle$ at fixed separation $r=10^{-3}L$. Changing $\Rey$
while keeping the system size $L$ the same changes the Kolmogorov length $\eta$ and hence
$R=r/\eta$ is different for different value of $\Rey$. Unless $R< \Rc$ (whether
this condition is satisfied or not is determined by the values of the Stokes numbers), the
relative velocity statistics depends on $R$, as the dissipation-range
theory shows. Thus evaluating the moments at $r=10^{-3}L$ for changing $\eta$ may give rise to a spurious
Re dependence.  It would be of interest to test quantitatively whether
the ${\rm Re}$-dependence predicted by the dissipation-range theory is consistent with this explanation.

It is a strength of the dissipation-range theory summarized in Section \ref{sec:theory} that it predicts
how the moments of relative velocities depend upon particle separation $R$. The microscopic dust grains
in accretion disks are much smaller than the Kolmogorov length $\eta$, so that the collision radius
$R=a_1+a_2$ is well below $\eta$.
 Inertial-range
theories \cite{volk1980collisions,Mizuno88,mar+miz+volk91,meh+usk+wil07,Gus08b} do not refer
to scales below $\eta$. As a consequence they cannot describe collisions that occur deep
in the dissipation range.  In DNS it is also difficult to reach to such small scales, 
much smaller than $\eta$, simply because particles rarely come so close. But collisional aggregation in turbulent aerosols is fluctuation dominated when the systems are dilute, so that such rare events matter. 
Several recent works~\cite{pan2013turbulence,pan2014turbulence, ishihara18} give
results for RMS relative velocities at fixed separations, usually of order $\eta$,
 irrespective of the size of the particles. The theory (\ref{eq:mp1}-\ref{eq:cpp}) allows 
 to extrapolate the DNS results to $R=a_1+a_2$. Here the parameter $\Rc=\Vc/z^\ast$
 plays an important role. If $R< \Rc$ then the theory shows that the relative particle-velocity statistics
 is independent of the separation $R$. 
  
 A weakness of the dissipation-range theory is that it expresses the prefactors
 $b_p$ and $c_p$ in the $R$-dependence of the moments in terms of
  parameters $z^\ast$, $\muc$, $\Vc$, and $V_0$ that must be determined separately, by DNS for example.
  The theory shows, moreover, that the prefactors are not universal. It would therefore be of great interest
  to find alternative ways of computing these prefactors. One possibility, although numerical, is to
  use the approach of Zaichik and collaborators \cite{zai+sim+ali03,Zai+ali+sin08} and
  its refinements \cite{pan2010relative}. 
    
\section{Summary and Conclusions}
\label{discussion}
Let us summarize the key findings here. We used direct numerical
simulations of particle-laden, homogeneous and isotropic, forced
turbulence to study the statistics of relative velocities and
separations between particles with different Stokes numbers.
We computed the joint  distribution 
of particle separations and their relative velocities.
 We found that the shape of the distribution is in good agreement
 with the predictions of dissipation-range theory 
\cite{meibohm2017relative}. When the
 difference between the two Stokes numbers is small enough,
 then the distribution exhibits power laws, and the exponent is
 related to fractal patterns in phase space \cite{gus+meh16}.
 We found that the power laws are cut off at small
 relative velocities, at a scale $\Vc$. We found that $\Vc$ depends
 linearly on $\theta$ for small values of $\theta$, 
in agreement with the theoretical prediction 
\cite{meibohm2017relative}.
 
 When $\theta$ is large, by contrast, theory predicts that the body of the distribution is broad Gaussian \cite{meibohm2017relative}, in agreement with the DNS of \cite{pan2014turbulence,ishihara18}.  In a log-log plot 
 Fig,~\ref{fig:jpdf} this Gaussian appears as a region where 
 $\mP/R^2$ is roughly constant. 
 The shape of the distribution beyond $V_0$ (here simply set to zero) is not known. 
There are indications  \cite{pan2014turbulence} that the theory
of Ref.~\cite{Gus08b} may work for the tails. But this could not be
unequivocally shown, and it must be borne in mind that the prediction of Ref.~\cite{Gus08b} applies to
large Stokes numbers in systems with a very well developed inertial range, so that the scale-dependent Stokes number
at the largest scale is much less than unity. These questions remain for further studies.

Dissipation-range theory  \cite{gustavsson2011distribution,Gus12,gustavsson2014relative,gustavsson2016collisions,meibohm2017relative} predicts how the relative-velocity fluctuations depend on particle separation. 
This power-law dependence of the relative-velocity moments upon particle separation is universal
(but the prefactors of the power laws are not). 
The  original inertial-range theories discussed above do not refer to particle  separations in the dissipation range, 
and attempts to modify inertial-range theories to take into account dissipation-range dynamics~\cite{ormel2007closed,pan2015turbulence}  were shown to fail (Fig.~5 in Ref.~\cite{ishihara18}),
so that they cannot be used to model collision velocities of microscopic dust grains in circumstellar accretion disks,
where collisions happen in the dissipation range. 
It is challenging to use DNS to determine collision rates and velocities of small grains deep in the dissipation range, because 
such encounters are infrequent, yet significant. Usually, DNS data on relative-particle velocities \cite{pan2013turbulence,pan2014turbulence, ishihara18} are evaluated
at fixed separations of order $\eta$, as discussed above. The theory described and tested here 
allows to extrapolate the DNS results to the relevant scales, often much smaller than the Kolmogorov length $\eta$. 

Note that \Eq{eq:vrms} is essentially an expansion
in powers of ${\rm St_2}$  for small ${\rm St_2}$ 
where we have retained terms up to
first order in ${\rm St_2}$. We have checked from our DNS
that the correlation function on the second lines of \Eq{eq:vrms}
is always negative and is proportional to ${\rm St_1}^2$ for small
${\rm St_1}$. \Eq{eq:vrms}, which is confirmed by our DNS, (\Fig{fig:moments} (c)),  
is clearly in disagreement with     
Abrahamson's theory ~\cite{abr75} which predicts that the
rms relative velocity of two inertial particles is given by 
the sum of their individual rms velocities.
This disagreement becomes apparent if we 
take the limit ${\rm St_2} \to 0$ in Eq. (18) in which
case the rms relative velocity appears as 
\textit{difference} between the rms velocities
of an inertial particle and a tracer. 
This is because Ref. ~\cite{abr75} assumes that the motion of the two
particles are uncorrelated -- an approximation
of dubious validity when the particles are close to each 
other, i.e., about to collide.    
This again illustrates one of the central messages
of this paper:  a theory of relative velocity
of two particles must take into account the distance
between them, otherwise the theory will fail
to predict collision velocities.

\section{Acknowledgments}
This work is supported by the grant Bottlenecks
for particle growth in turbulent aerosols from
the Knut and Alice Wallenberg Foundation (Dnr. KAW
2014.0048), by  Vetenskapsradet [grants 2013-3992 and 2017-03865], and Formas [grant number 2014-585].
 The computations
were performed on resources provided by the
Swedish National Infrastructure for Computing (SNIC)
at PDC. DM and AB thank John Wettlaufer for useful discussions.


\end{document}